# The elusive nature of roughness: linking hydraulics and graph theory for water distribution networks model calibration

Karol Dykiert[1,2,3*], Mateusz Stolarski[3**], Michał Czuba[3***], Wojciech Cieżak[2****], Piotr Bródka[3*****]

[1]MPWiK S.A. (Municipal Water and Sewerage Company) in Wrocław, Na Grobli 19, Wrocław, Poland

[2]Department of Water and Wastewater Management and Waste Technology, Wrocław University of Science and Technology, Wybrzeże Stanisława Wyspiańskiego 27, Wrocław, Poland

[3]Department of Artificial Intelligence, Wrocław University of Science and Technology, Wybrzeże Stanisława Wyspiańskiego 27, Wrocław, Poland

[*]_corresponding author: karol.dykiert@pwr.edu.pl_, [**]_mateusz.stolarski@pwr.edu.pl_, [***]_michal.czuba@pwr.edu.pl_, [****]_wojciech.ciezak@pwr.edu.pl_, [*****]_piotr.brodka@pwr.edu.pl_

**Keywords:** Pipe roughness, Water distribution network, hydraulic modelling, Computational network science, Clustering, Evolutionary algorithm

**Abstract:** Accurate pipe roughness estimation in large-scale water distribution networks is often hindered by the high cost of traditional field methods. This study investigates whether network partitioning, by utilizing hydraulic and graph-derived attributes, can enhance the calibration of these parameters. Using a high-fidelity model of a real network as a benchmark, we evaluate density-based clustering, and topology-driven grouping strategies. Optimization experiments demonstrate that attribute-based grouping yields stable, repeatable results comparable to manual calibration for hydraulically significant pipes. While hydraulic attributes generate more distinct cluster structures, the inclusion of graph-based data improves calibration robustness by stabilizing the optimization process. Notably, density-based clustering achieves similar accuracy to k-means while reducing computational effort in specific configurations. Although the method does not eliminate all sources of uncertainty, results suggest that topology-informed grouping provides a systematic, reproducible, and computationally efficient alternative to manual heuristics, highlighting the critical role of network structure in reliable parameter estimation.

# 1. Introduction

## 1.1. The problem of roughness formulation

Water resources have become increasingly critical under the pressures of climate change, which exacerbates hydrological extremes such as prolonged droughts, increased evaporation, and altered precipitation patterns [1]. These phenomena place additional stress on drinking water supply systems, intensifying the need for efficient, resilient, and well-managed water distribution infrastructure. In response to these challenges, emerging policies and recommendations aim to strengthen regulatory and strategic frameworks for drinking water supply, with particular emphasis on reducing water losses and safeguarding water quality, where non-revenue water reduction, infrastructure renewal, and improved monitoring are consistently highlighted in both policy initiatives and scientific recommendations as key measures for long-term sustainability and public health protection [2–4].

Pipe roughness is a significant contributing factor to both water quality deterioration and increased water and energy losses in distribution networks. Accurate estimation of pipe roughness, however, remains a challenging task. The most robust and widely accepted approach relies on hydrant discharge tests, yet this method is inherently cumbersome and time-consuming, particularly for large-diameter pipes within extensively looped distribution networks [5,6]. Under such conditions, reliable roughness assessment often requires multiple valve shut-offs and substantial discharge volumes to establish measurable hydraulic gradients. Our heuristics follow others in the idea that more efficient alternatives can be developed, ones which rely on various computational methods.

## 1.2. Computational methods of roughness calibration

While field-based discharge tests remain the benchmark for accuracy, the transition to computational calibration introduces several numerical challenges. The reliability of these models depends on various factors beyond friction alone, such as the precision of nodal elevation data [7] or the strategic placement of pressure sensors [8]. However, even with accurate boundary conditions, the high dimensionality of large-scale water distribution networks (WDNs) makes individual pipe estimation an ill-posed problem where many different roughness combinations can yield the same pressure results. To address this, researchers have developed various strategies ranging from rigorous mathematical inversion to heuristic grouping.

To manage the lack of data for every individual pipe, Kumar et al. focused on state and parameter estimation frameworks that utilise graph-theoretic concepts to reduce dimensionality [9]. Their approach employs a systematic grouping procedure (using the K-means clustering algorithm) to cluster pipes with similar physical characteristics, such as material or age, into common calibration zones. This allows the model to estimate a shared roughness coefficient for the group, significantly increasing computational efficiency and providing a better estimate of the network state even when some demands are not fully measured. This concept was further evolved through the use of nonlinear state observers, which treat the network as a system of damped oscillators to allow for continuous, online calibration whenever real-time flow data is accessible [10].

However, the pursuit of high-resolution calibration can lead to "overfitting", where the model captures data noise rather than the physical reality. Recognising this, Zhao et al. advocated for a "simpler is better" philosophy [11]. Their comparison of various optimisation strategies, including evolutionary algorithms and gradient-based methods, demonstrated that a finer discretisation of the search space does not necessarily guarantee better results. Instead, they found that coarser, more stable pipe groupings often lead to more robust calibration in real-world scenarios.

Yet the challenge remains in determining exactly how these attributes should be utilised to achieve optimal partitioning. Understanding the effectiveness of these groupings requires a deeper investigation into the structural properties of the network itself. This necessitates a shift from purely hydraulic adjustments to a broader perspective that treats the pipe network as a mathematical graph.

## 1.3. Bridging graph-based methods with environmental engineering

A WDN is a network and can therefore be modelled as a graph, that is, a structure composed of vertices and edges connecting pairs of them [12]. Unfortunately, the graph underlying such a system is usually considered by engineers only within a spatiotemporal context (e.g., during design, the network must avoid collisions with existing infrastructure). By contrast, electronics – a field ostensibly distant from environmental engineering – exploits more advanced graph-based methods in circuit theory, a domain conceptually close to the problem addressed in this work. In particular, Kirchhoff's laws make use of spanning trees to derive equivalent resistances of electrical circuits, an analogous quantity to the equivalent roughness of a pipeline [13]. This established practice legitimises the benefits of adopting a graph-oriented perspective into non-obviously related technical problems.

As this study examines whether integrating network science with environmental engineering improves estimates of WDN equivalent roughness, three core graph concepts are briefly outlined. First, the construction of a graph representation suitable for roughness calibration, as the parameter of interest is associated with links rather than nodes and different network regions may carry unequal hydraulic importance; this motivates the use of graph coarsening, which reduces network complexity while preserving essential structural information [14]. A related challenge concerns graph clustering – commonly expressed through community detection – assigns nodes to groups that are more densely interconnected internally than with the rest of the network [15]. Finally, centrality measures quantify the relative importance of network elements, with metrics such as betweenness indicating structural control over flow pathways [16]; in WDN analysis, such measures can help identify components whose structural role disproportionately influences system behaviour and calibration outcomes.

### 1.4. The focus of our study

In this work, we synthesise Kumar's [9] grouping logic with the reduced-complexity requirements advocated by Zhao [11]. Building on our previous study [17], which established a dataset of hydraulic and graph-based attributes to mitigate the intrinsic bias of manual pipe grouping, the present research seeks to determine the optimal configuration of this framework. Using a high-fidelity hydraulic model as a benchmark, we examine the extent to which different network partitioning approaches can enhance calibration performance. Rather than proposing a purely theoretical method, the study provides a systematic evaluation of how grouping logic and attribute selection influence pipe roughness calibration, with the aim of identifying efficient pathways to improved model accuracy in complex looped systems.

### 1.5. Article structure

This paper is structured as follows. **Materials and Methods** introduce the study area, dataset, methodological framework of pipe grouping, and the application of the Shuffled Complex Evolution (SCE) [18] algorithm for roughness calibration. The **Results** section presents the clustering analysis, evaluates multiple grouping strategies using established clustering metrics with secondary descriptors, and assesses their impact on calibration performance along with the accuracy of roughness estimation. The **Discussion** focuses on grouping and clustering behaviour, quantitative analysis of grouped pipes, optimisation repeatability, and the influence of these stages on pipe roughness. Finally, the **Conclusions** part summarises the key findings, outline the practical implications for roughness calibration in WDNs, and identify directions for future research.

The simplified workflow, describing the research steps, is shown in Figure 2.

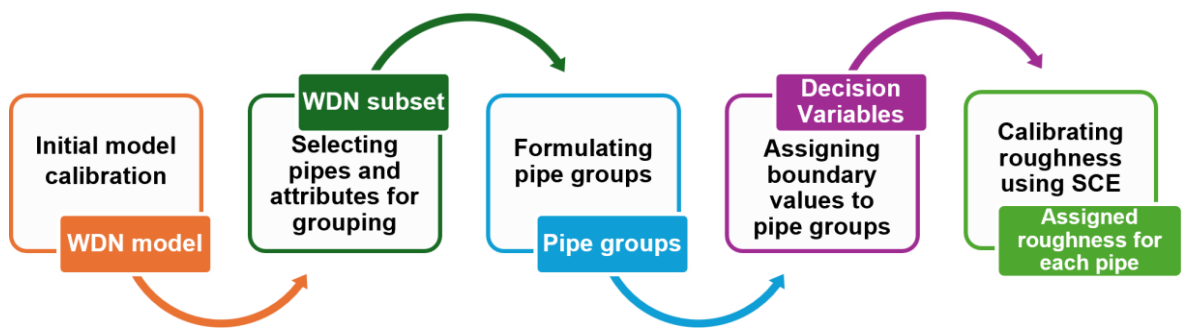

*Figure 1. Research workflow.*

## 2. Materials and Methods

The analysed system is a high-pressure zone (HPZ) of the WDN (Figure 2), consisting of six District Meter Areas (DMAs). The dataset, associated preprocessing procedures and the initial model calibration were described in detail in our previous publication [19]; concise summaries are provided in Appendices A and B for reference.

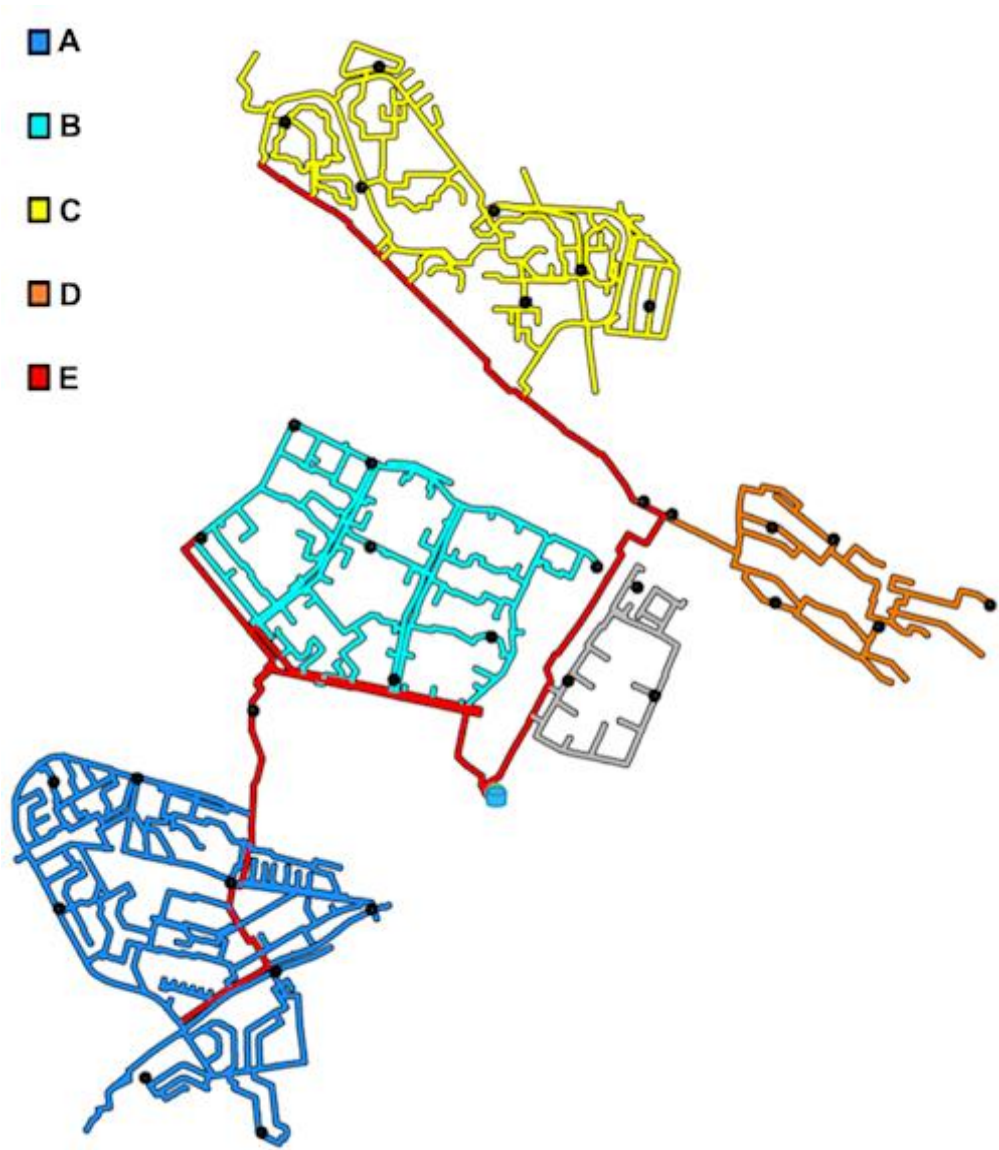

*Figure 2. Hydraulic model of HPZ in the city of Wrocław. Pipe colours denote DMA assignments; grey DMA was excluded from the final calibration stage (see 3. Results). Pressure measurement stations are marked with black dots.*

To avoid ambiguity arising from the multi-stage nature of the proposed workflow, we clarify the terminology used throughout this article. The term **grouping** is used broadly to refer to the partitioning of network elements into groups and encompasses classical clustering approaches (e.g., k-means, HDBSCAN), as well as network coarsening and community detection. The resulting pipe **clusters**, **super-nodes**, and **communities** are collectively referred to as **groups**.

These groups serve as **Decision Variables (DVs)** for the SCE roughness optimisation, with lower and upper bounds for acceptable roughness values assigned as specified in the Methodology section. The term optimisation is reserved exclusively for the SCE roughness calibration stage. For all roughness calibration in this study, the **Darcy–Weisbach formulation** was employed.

## 2.1. Selecting pipes and attributes for grouping

The clustering dataset consisted of all pipes selected for roughness calibration, together with their associated attributes. Those included fundamental physical and categorical attributes as well as flow rate, derived from the initial simulation. Regarding graph-based attributes, commonly used descriptors were selected to characterise pipes (edges) and their topological relationships with junctions (nodes) [20]. This methodology was first described in [17] and remains unchanged for the current study. A consolidated list of all attributes used in the analysis is provided in Table 1.

*Table 1. WDN attributes selected for clustering. Reproduced from [17].*

| Hydraulic attributes | Graph attributes |
| --- | --- |
| Diameter | Bridge (edge-cut) |
| Length | Edge betweenness |
| Age | Degree sum |
| Role (transmission or distribution) | Degree difference |
| Material | Minimum degree |
| DMA assignment | Maximum degree |
| Flow rate (minimum, maximum, mean, median and standard deviation represented separately) | Edge strength |
| | Average neighbour degree difference |
| | Personalized PageRank similarity |

## 2.2. Formulating pipe groups

Pipe grouping was performed using the following algorithms: k-means clustering, HDBSCAN, Variation Neighbourhood (VN) coarsening, and Louvain community detection. For k-means and HDBSCAN, an additional clustering step was carried out using a reduced dataset containing only hydraulic attributes. Coarsening and community detection were applied directly to the HPZ network subset.

**K-means** [21] clustering is a centroid-based, unsupervised learning algorithm that partitions the data into k clusters by minimising the within-cluster sum of squared distances (inertia). The algorithm iteratively assigns observations to the nearest cluster centroid and updates centroid positions until it converges. The primary hyperparameter, the number of clusters k, was selected using the elbow method, a commonly applied empirical approach [22].

**HDBSCAN** [23] is a density-based clustering method that identifies clusters as regions of higher data density and explicitly classifies low-density observations as noise, "-1" cluster. The algorithm is capable of detecting clusters of varying shape and density without requiring the number of clusters to be specified. In this study, all HDBSCAN hyperparameters were kept at their default values, following the recommendations in [24]. To the best extent of our knowledge, HDBSCAN has not yet been tested as a preprocessing model of the water supply pipeline in the task of roughness estimation. However, the very different principles guiding the algorithm offer potentially interesting alternatives to k-means.

Both clustering approaches were applied to identify characteristic patterns in the data and to support the selection of the physical and operational features relevant to the pipe roughness calibration process.

As the dataset contains numerous additional attributes that may be difficult to obtain, particularly due to the level of graph-theoretical knowledge required, the process of clustering was performed using a reduced dataset comprising only hydraulic attributes. For HDBSCAN, all hyperparameters were retained at their default values. For k-means, the elbow-based optimisation of the number of clusters was not repeated; instead, the number of clusters was fixed at **k = 27**, consistent with the original study [17], as this value was previously shown to yield satisfactory results.

**Coarsening** is a type of graph reduction method that creates a smaller graph while preserving its essential structural information [25]. During this process, the original nodes are grouped into super-nodes, and edges between these groups are aggregated into super-edges according to a predefined aggregation scheme [26]. In this study, we employ the VN method, which merges nodes with similar neighbourhood connectivity patterns, following [14]. Consequently, the coarsened graph maintains key local connectivity characteristics of the original graph [25]. Due to the nature of this task in the calibration pipeline – namely, preprocessing the graph edges for the genetic algorithm – it was necessary to apply a line graph transformation. To improve the computational efficiency of the genetic algorithm used for pipe roughness estimation, we assume that the coarsened graph should contain the minimum feasible number of super-nodes. Consequently, the VN reduction rate was set close to 1, and a post-processing step was applied to handle singleton super-nodes by merging them with neighbouring groups. The resulting behaviour is conceptually analogous, though not algorithmically equivalent, to HDBSCAN, where points assigned to the label "-1" are treated as noise.

**Community detection** was employed as a second alternative to classical clustering algorithms, to capture the intrinsic community structure of real-world networks – a fundamental feature of graph models representing complex systems [27]. In this work, we used the Louvain community detection algorithm, a widely used unsupervised method that extracts the community structure of networks by optimising modularity [28]. The Louvain algorithm proceeds iteratively in two main phases. First, each node is assigned to its own community, then the nodes are reallocated to neighbouring communities if such moves increase modularity. After optimisation of local modularity, the communities are aggregated into a new network, with each community treated as a single node. This process is repeated until no further improvement in modularity is observed. For directed networks, modularity is computed considering both in-degree and out-degree, allowing the algorithm to account for edge directionality [29]. The assumptions regarding coarsening were applied in this approach as well, including minimising the number of communities by lowering the resolution parameter and marking those containing single nodes as noise.

## 2.3. Assigning boundary values to pipe groups

To complete the formulation of DVs for the SCE-based roughness optimisation, lower and upper bounds were assigned to each pipe group based on the smallest diameter within the group. These bounds defined the admissible range of roughness values from which the SCE algorithm could sample. The minimum roughness value was set equal to **0.2** times the smallest pipe diameter within a group, while the maximum value was set to **0.75**, or **0.3** in the case of groups consisting exclusively of transmission mains. These bounds were selected based on engineering judgement and previous research [17,19].

## 2.4. Calibrating roughness using Shuffled Complex Evolution

Roughness calibration was carried out using the SCE global optimisation algorithm [18], which is embedded in the modelling software employed in this study and was therefore selected to ensure methodological consistency and reproducibility. SCE combines concepts from evolutionary optimisation and the simplex method by evolving multiple complexes in parallel, periodically shuffling information between them to balance global exploration and local convergence.

Hyperparameter values were defined following the recommendations of the original SCE formulation [18]. To ensure comparability across optimisation experiments, all hyperparameters were kept fixed throughout the study. Pipes roughness calibration was performed using pressure measurements as reference data, and the objective function was defined as the **Root Mean Square Error** (RMSE) between simulated and observed pressures. The target function can be described as:

$$\min\left(\sum_{m=1}^{M} RMSE_m \cdot \omega_m\right) \quad (1)$$

Where **$RMSE_m$** is RMSE calculated for **m-th** pressure measurement point and **$\omega_m$** is weight assigned to **$RMSE_m$**. Weights were generally set to 1, but for six pressure series they were adjusted based on field expertise to account for minor leaks or uncertain node elevations, which could not be fully corrected in the hydraulic model, and the measured data were not considered entirely erroneous, which is why weight adjustments were applied rather than data exclusion.

The use of RMSE was deemed sufficient based on its demonstrated effectiveness in previous studies [17]. A complete set of SCE's hyperparameters and their values is provided in Appendix C.

## 2.5. Metrics and supportive descriptors

Obtained results have been analysed in three distinct steps, which are presented in Figure 3.

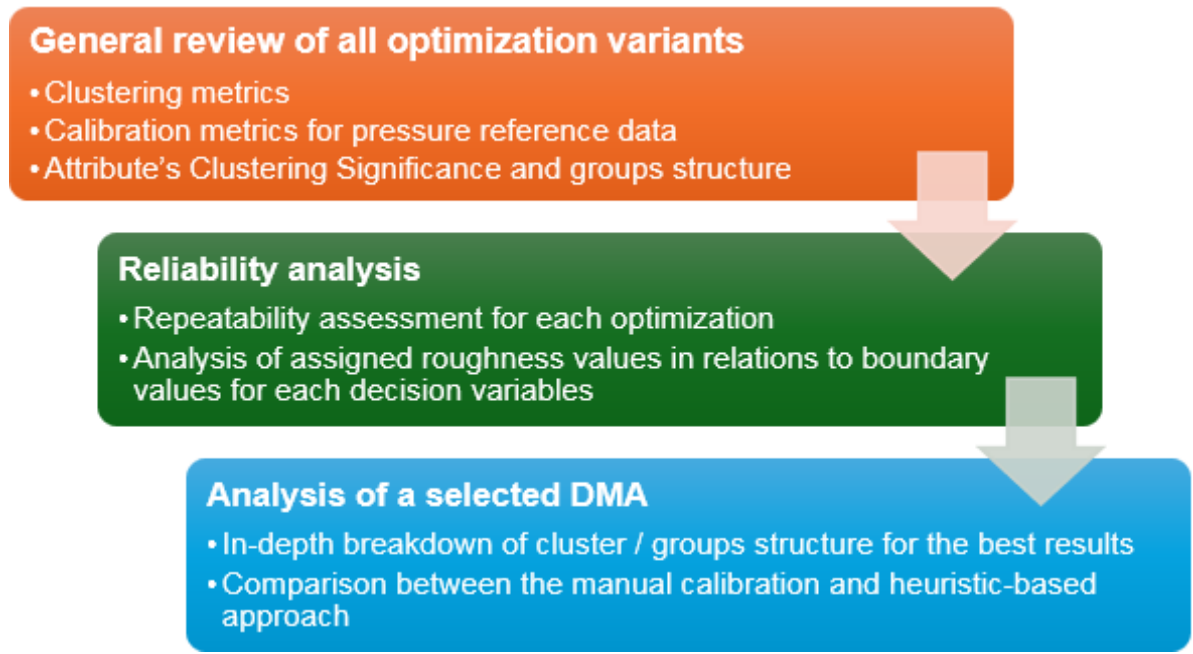

*Figure 3. Results analysis summary.*

Clustering performance was evaluated using three internal validation metrics: the **Silhouette Coefficient (SC)** [24], the **Davies-Bouldin Index (DBI)** [24], and the **Calinski-Harabasz Index (CHI)** [24]. For the k-means clustering, inertia were additionally computed to support the elbow method for selecting the optimal number of clusters.

Following roughness calibration using the SCE algorithm, we evaluated model performance by computing the mean **RMSE** and the **Index of Agreement (IOA)** between measured and simulated pressures across all pressure reference points in the network. In addition, the average computational time required by the SCE algorithm was recorded for each optimisation run.

To support the assessment, a set of custom descriptors was formulated based on fundamental mathematical concepts, such as the mode and variance. These descriptors do not introduce new statistical metrics but adapt established descriptive measures to the specific requirements of the optimisation and clustering analyses performed in this study.

To examine the contribution of individual attributes to the grouping process, the **Attribute's Clustering Significance (ACS)** was computed, consistent with previous studies [17]. ACS quantifies an attribute's influence on cluster formation by counting the number of instances in which within-group normalised variance falls below a specified threshold, indicating a strong contribution to group definition:

$$ACS = \frac{1}{K}\sum_{k=1}^{K} 1_{\left\{\frac{s_k^2}{\max_a(s_a^2)} \leq 0,1\right\}} \quad (2)$$

Where **K** is the total number of clusters, $s_k^2$ is the attribute's variance within cluster **k**, and $\max(s_a^2)$ is the maximum variance observed among all **a** attributes across all clusters.

To assess the behaviour of DVs during the roughness optimisation, the proportion of cases in which a DV reached its prescribed boundary value was calculated and normalised. This descriptor is referred to as the **Boundary Index (BI)**:

$$BI = \frac{1}{N \cdot K}\sum_{n=1}^{N}\sum_{k=1}^{K} f(b_{nk}) \quad (3)$$

Where $f(b_{nk})$ equals 1 if the calibrated roughness of the **k-th** DV (corresponding to the **k-th** cluster in Eq. 2) reaches either its upper or lower boundary, and 0 otherwise. The sum is averaged over **N** optimisation runs. **BI = 0** indicates no DV reached a boundary, while a value of **BI = 1** indicates all DVs did.

Optimisation repeatability was quantified using the **Repeatability Index (RI)**, which measures the consistency of calibrated roughness:

$$RI = \frac{1}{K}\sum_{k=1}^{K} \frac{g(\varepsilon_k) - 1}{N - 1} \quad (4)$$

Where $g(\varepsilon_k)$ denotes the number of times the modal calibrated roughness values of the **k-th** DV occurs across all optimization runs. When calculating $g(\varepsilon_k)$ we allow for 10% margin of error, reflecting minor acceptable deviations. RI ranges from 0 to 1, with 1 indicating complete consistency and 0 indicating the complete lack of repeatability.

Each optimisation experiment was repeated five times to ensure the robustness of the dataset and the stability of the results.

Finally, pipe roughness values obtained using the heuristic were compared with those derived from manual calibration for DMA D (Figure 2). To quantify differences at the pipe level, the **Relative Absolute Error (RAE)** was calculated, and the **Mean Absolute Percentage Error (MAPE)** was subsequently computed as an overall indicator across all analysed pipes.

Technical data regarding utilized software and hardware are included in Appendix D.

# 3. Results

This section presents the outcomes of the clustering and calibration experiments. The results are reported in terms of clustering quality, optimisation performance, and the consistency of calibrated roughness values.

## 3.1. K-means and the elbow method

To determine the optimal number of clusters for the k-means algorithm, we employed the elbow method. The principal criteria for this selection were inertia and the number of clusters, supplemented by the SC and the DBI to support the decision. The CHI was not considered, as its variation across the tested cluster numbers was negligible.

Cluster counts were evaluated in increments of two, ranging from **11** to **99**, with **k = 27** serving as a reference point based on our prior work. The results of the elbow analysis are shown in Figure 4. The objective was to identify a cluster count that provided the most favourable trade-off between clustering performance and computational efficiency.

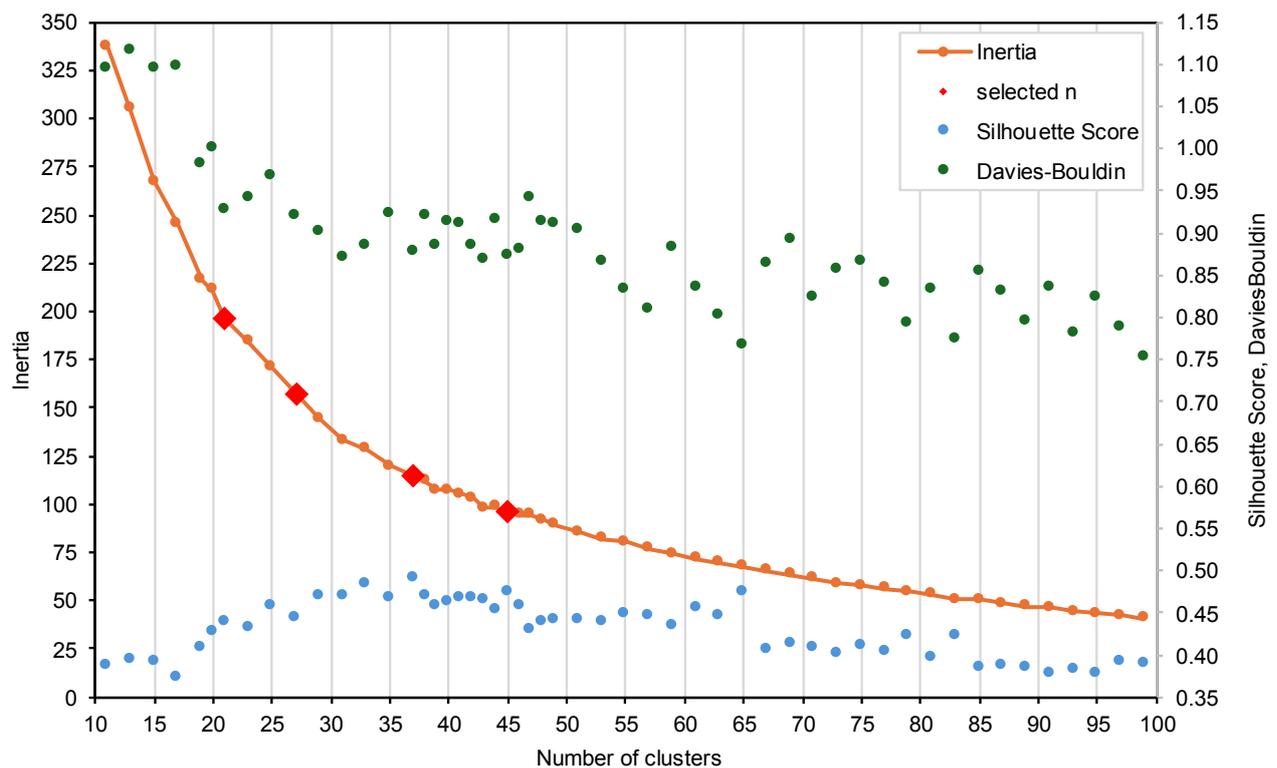

*Figure 4. Identification of the optimal k in k-means with the elbow method.*

Both inertia and the DBI decreased as the number of clusters increased, indicating improved cluster compactness and separation. The SC initially rose with increasing k, but began to decline slightly around **k = 27**, suggesting that clustering quality improved up to that point and then gradually deteriorated.

Because higher numbers of clusters substantially increase computational demand, we examined the region around **k = 45** in greater detail. In this interval, cluster numbers were evaluated with a step of one to ensure the identification of a local optimum. Differences between adjacent cluster counts remained small for both the SC and the DBI.

Based on this analysis, four cluster counts were selected for subsequent roughness optimisation:

- **45** (**CL_k45**) – the largest tested cluster count, yielding the overall best clustering metrics,
- **37** (**CL_k37**) – the second-best option, with a slightly higher SC but only marginally worse inertia and Davies-Bouldin values,
- **27** (**CL_k27**) – the baseline cluster count used in earlier research,
- **21** (**CL_k21**) – the smallest cluster count that still produced metrics comparable to the selected alternatives.

### 3.2. Additional grouping methods

As described in the Methodology section, we additionally applied k-means clustering to the dataset with graph-theory-derived attributes removed. For consistency with our previous work, the number of clusters was fixed at **k = 27** (**CL_k27_NG**).

For the HDBSCAN algorithm, clustering was performed using the default hyperparameter settings provided by the scikit-learn implementation. The algorithm was applied to both the complete subset (**CL_HDB**) and the subset without graph-based attributes (**CL_HDB_NG**).

In addition, we explored topology-based grouping approaches by clustering nodes using the VN method (**VarNg**) and the Louvain community detection algorithm (**LouvainCom**).

All model performance metrics, as well as clustering evaluation metrics (where applicable), are presented in Table 2, together with the number of DVs / groups used in each optimisation case. The "-1" outlier group was not included in that number.

Table 2. Complete list of grouping and calibration parameters.

| Optimisation variant ID | No. of DVs | SC | DBI | CHI | inertia | mean RMSE, m $H_2O$ | mean IOA | mean SCE roughness optimisation time, min |
|---|---|---|---|---|---|---|---|---|
| **CL_k21** | 21 | 0.44 | 0.93 | 199.45 | 196.02 | 0.33 | 0.62 | 27.6 |
| **CL_k27** | 27 | 0.44 | 0.92 | 195.82 | 157.36 | 0.33 | 0.62 | **24.3** |
| **CL_k37** | 37 | **0.49** | 0.88 | 197.46 | **115.01** | 0.33 | 0.62 | 52.8 |
| **CL_k45** | 45 | 0.48 | 0.87 | 194.79 | **95.72** | **0.32** | **0.63** | 57.6 |
| **CL_HDB** | 40 | **0.51** | 0.71 | 224.35 | NA | 0.33 | **0.63** | 27.9 |
| **LouvainCom** | 19 | NA | NA | NA | NA | 0.37 | 0.57 | **18.4** |
| **VarNg** | 35 | NA | NA | NA | NA | 0.36 | 0.56 | 33.4 |
| **CL_k27_NG** | 27 | 0.46 | **0.62** | 1285.05 | 19.20 | **0.32** | 0.64 | 22.1 |
| **CL_HDB_NG** | 25 | **0.54** | **0.61** | 1054.64 | NA | **0.32** | **0.64** | 28.6 |

Several general patterns can be identified. HDBSCAN produced slightly better clustering quality overall. For k-means, increasing k improved SC and DBI, while CHI remained largely unchanged. The most pronounced differences were observed between clustering the full attribute set and the reduced, graph-free subset, with the reduced subset yielding better clustering metrics, particularly higher CHI and lower inertia in the k-means variants. These differences, however, did not translate into substantial changes in calibration performance, as indicated by RMSE and IOA, which remained comparable across optimizations variants.

### 3.3. Attributes significance analysis

To provide further insight into the clustering results, we computed the ACS to quantify which attributes contributed most to cluster formation in terms of within-group variance. For comparison, ACS values were calculated for the k-means and CL_HDB variants, as well as for LouvainCom and VarNg.

Across all clustering methods, hydraulic attributes consistently exhibited low within-group variance, resulting in high ACS scores, indicating a strong influence on group formation. Attributes that consistently achieved high ACS across variants included role, DMA, and is_bridge (edgecut). Among graph-related features, minimum and maximum node degree generally displayed high ACS values.

For k-means clustering, ACS scores for graph attributes tended to increase with the number of clusters, whereas CL_HDB achieved the highest overall ACS values when considering both hydraulic and graph-based attributes. LouvainCom and VarNg performed well in terms of graph attributes but remained below CL_HDB.

Direct comparison across clustering variants is limited by the differing number of attributes; nevertheless, variants based on the full attribute subset consistently achieved higher ACS than those relying solely on hydraulic attributes. These results, summarised in Figure 5, provide a clear overview of which attributes primarily drive cluster formation, though they are supportive descriptors and do not directly indicate calibration performance.

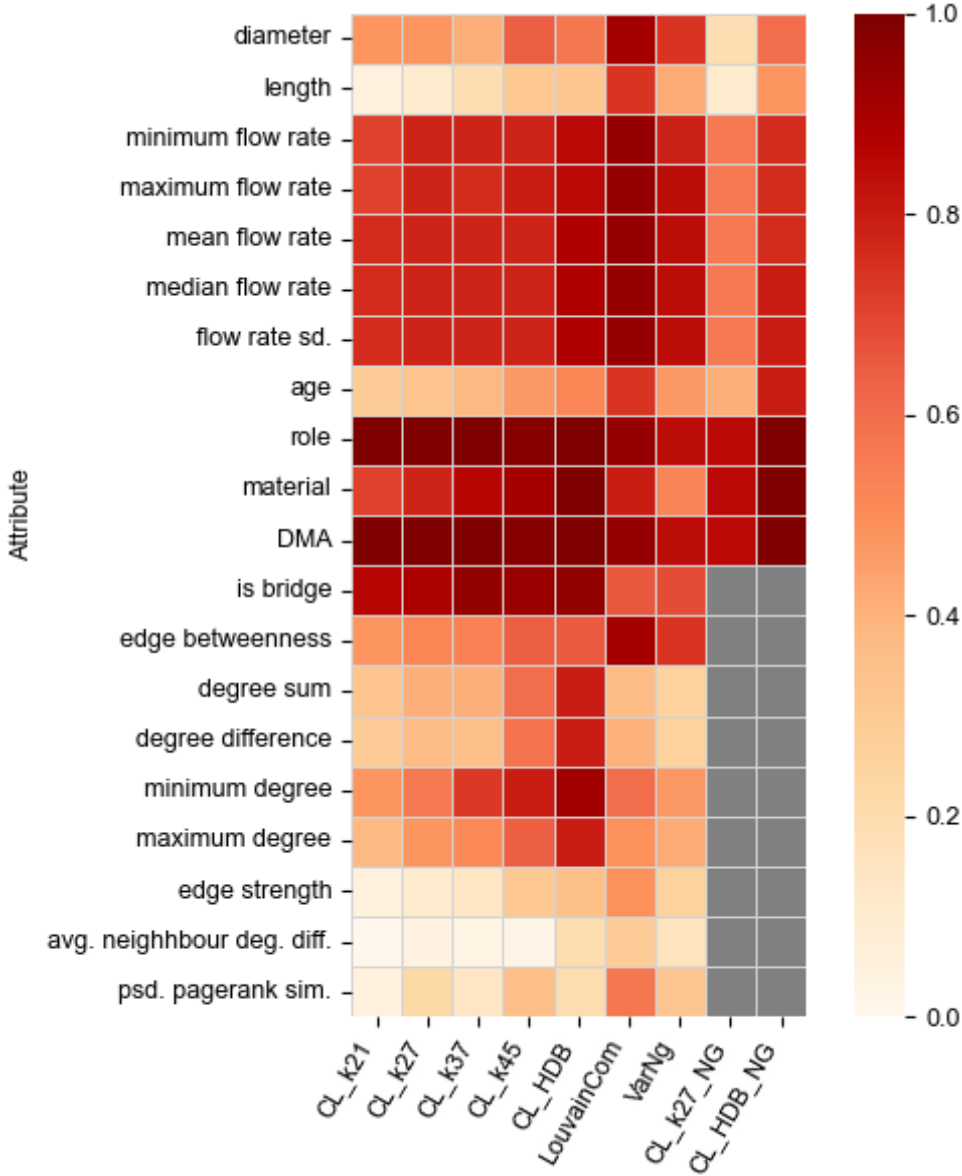

*Figure 5. ACS scores for all optimisation variants. Grey squares indicate missing values.*

## 3.4. Quantitative clustering analysis

To complement the ACS analysis, we conducted a quantitative examination of cluster structures across all optimisation variants using simple boxplots (Figure 6). One immediate observation is that both LouvainCom and VarNg exhibit a small number of groups containing a disproportionately large number of pipes. These groups, which are distinct from the "-1" clusters representing outliers (also marked in the figure), dominate the subset: while the majority of groups contain 2-25 pipes, several encompass more than half of the entire subset in total.

Since one of our goals was to achieve a well-diversified partitioning of the WDN, these clustering results are considered heavily biased and are therefore excluded from further analysis.

Similarly, for CL_k27_NG and CL_HDB_NG, the distribution of cluster sizes shows more irregularity than in their full-subset counterparts, although not as extreme as in the coarsening and community detection variants. The observed unequal distribution of clusters in CL_k27_NG aligns with its relatively lower ACS scores.

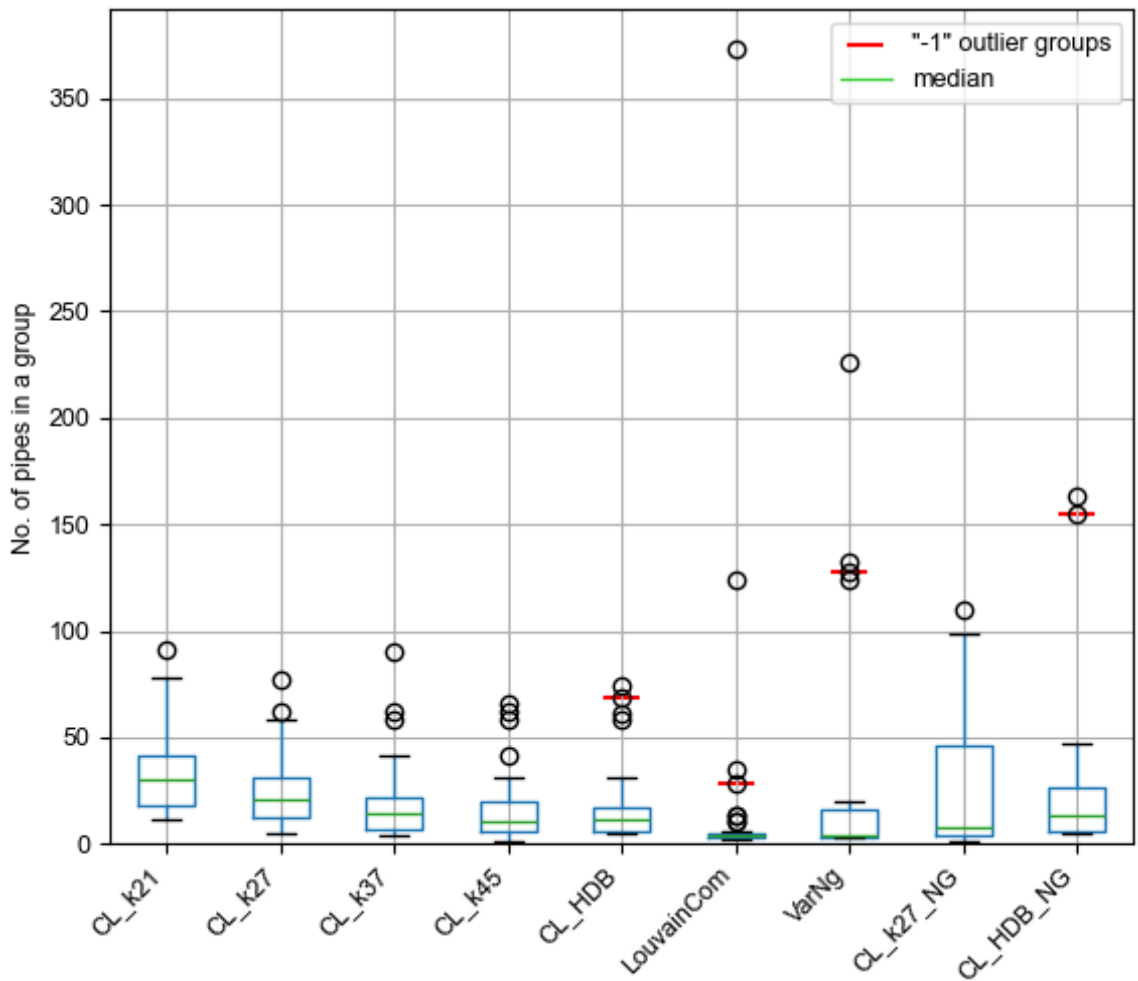

*Figure 6. Quantitative analysis of grouping results.*

### 3.5. Reliability analysis

Following the evaluation of grouping performance and its potential implications for pressure fit, we shifted the focus to roughness calibration itself, examining the behaviour of DVs across optimisation runs. To this end, the Boundary Index and Repeatability Index were used to assess, respectively, the frequency with which calibrated roughness values reached their prescribed bounds and the consistency of optimisation results across five runs (Table 3).

Among the remaining variants, CL_k45 exhibited the highest BI and was the only configuration to achieve a RI of 1, indicating fully repeatable outcomes across all runs. CL_k21 and CL_HDB also demonstrated high repeatability, whereas CL_k37 showed a noticeably lower RI despite a favourable BI. Based on RI, which was considered the primary criterion for robustness, CL_k45, CL_k21, and CL_HDB were deemed suitable for further analysis.

*Table 3. Summary of BI and RI for remaining optimisation variants.*

| Optimisation variant ID | BI | RI |
| --- | --- | --- |
| **CL21** | 0.53 | **0.87** |
| **CL27** | 0.53 | 0.72 |
| **CL37** | **0.39** | 0.73 |
| **CL45** | 0.36 | **1.00** |
| **HDBSCAN** | **0.41** | **0.83** |
| **CL27_NG** | 0.50 | 0.70 |
| **HDBSCAN_NG** | 0.46 | 0.58 |

## 3.6. Calibration comparison for a selected DMA

While it is cumbersome to track every individual change across the entire dataset, several consistent patterns can nevertheless be identified. To examine the structure of clusters in relation to roughness, we mapped the clusters assigned to pipes for the three best-performing optimisation variants (CL_k45, CL_HDB and CL_k21) for DMA D (Figure 2), used here as a representative example (Figure 7).

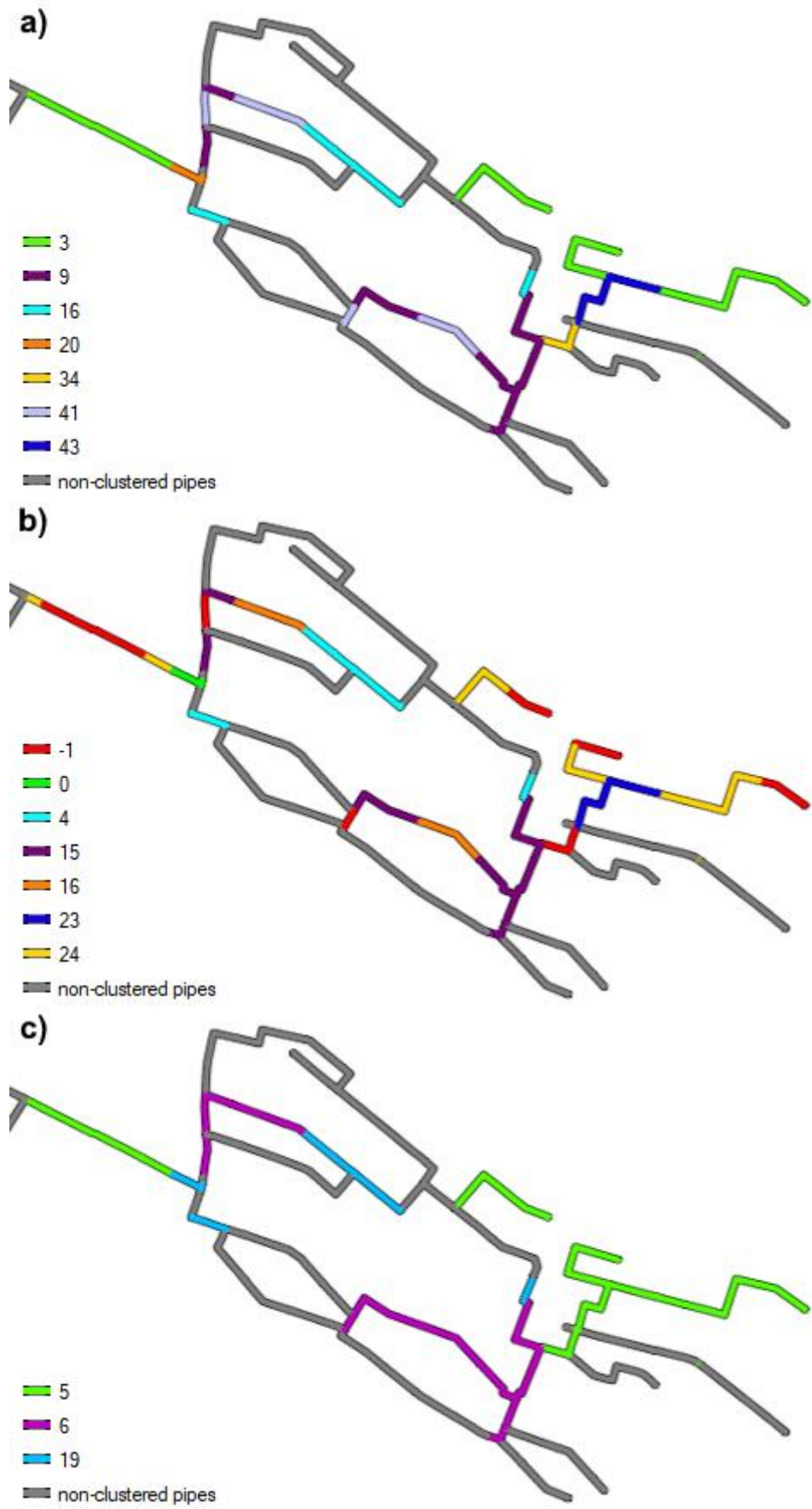

*Figure 7. Mapped clusters for a) CL_k45, b) CL_HDB, c) CL_k21.*

All clustering variants successfully identified and grouped branched elements of the network; however, in each case these groups also included the main transmission pipe located on the left side of the DMA (cluster no. 3 in CL_k45, corresponding to clusters no. "-1" and 24 in CL_HDB, and cluster no. 5 in CL_k21). Since CL_k21 contains substantially fewer clusters than CL_k45 or CL_HDB, direct one-to-one comparison is less straightforward. Based on visual inspection, CL_HDB and CL_k21 appear to further subdivide cluster structures identified in CL_k21, indicating a finer spatial resolution of grouping. A brief correspondence between cluster equivalents across the grouping variants is provided in Table 4. Finally, it is worth noting that, across all of them, clusters remain confined within the boundaries of the assigned DMA, except for the "-1" outlier cluster in CL_HDB, which extends beyond DMA limits.

*Table 4. Clusters similarity handout.*

| CL45 | HDBSCAN | CL21 |
|---|---|---|
| 9, 41 | -1, 15, 16 | 6 |
| 3, 34, 43 | -1, 23, 24 | 5 |
| 16, 20 | 0, 4 | 19 |

## 3.7. Comparing roughness values

As a final step in the results analysis, we compared the pipe-level roughness values obtained from all optimisation variants with those derived from manual calibration, which served as the reference. The comparison was again conducted for DMA D (Figure 2); to maintain clarity and relevance, only pipes with lengths of **50 m** or greater were considered. To quantify similarity, we calculated the Relative Absolute Error between roughness values obtained through the heuristic and those from manual calibration (Figure 8). Additionally, we computed the mean of these values, yielding the Mean Absolute Percentage Error.

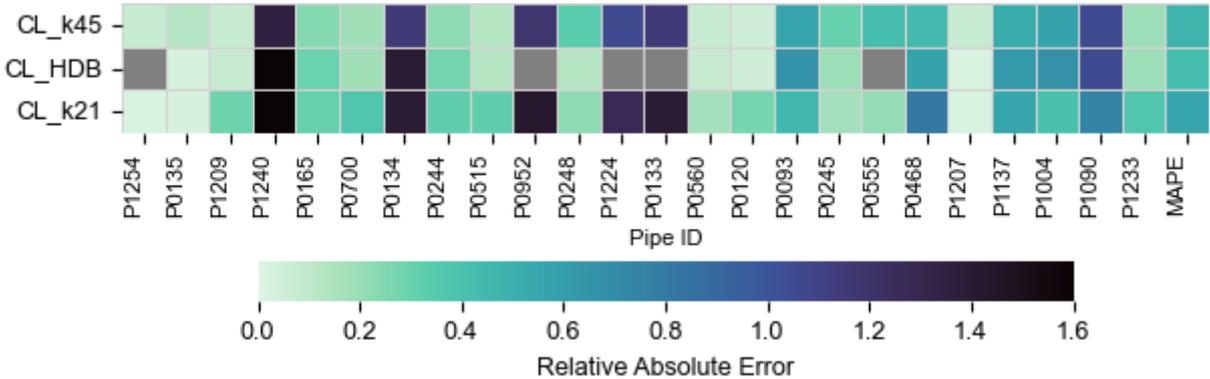

*Figure 8. Roughness similarities heatmap. Values closer to 0 indicate better similarity to values of roughness calibrated manually.*

The first notable observation is that roughness values remain generally consistent across all optimisation variants. Minor differences are observed for individual pipes (e.g., P1004 and P0165), and in several cases the SCE optimisation yielded roughness values that differed substantially from those obtained through manual calibration, although typically for the same pipes (e.g., P1240). Missing values in the CL_HDB variant are associated with the "-1" outlier cluster, which was excluded from the SCE optimisation; for these pipes, roughness values remained unchanged from the initial calibration. MAPE values were **0.48**, **0.43**, and **0.55** for CL_k21, CL_HDB, and CL_k45, respectively. Overall, differences between optimisation variants remain limited, and no substantial deviations are observed.

## 4. Discussion

This section interprets the observed patterns in clustering structure and calibration performance. The results are examined in the context of network grouping strategies and their implications for roughness estimation in water distribution networks.

### 4.1. Determining the number of clusters in k-means

The elbow method behaved largely as expected; however, it remains subject to well-known limitations, as it relies exclusively on inertia to identify the optimal number of clusters, which may result in arbitrary or biased decisions. In our case, the location of the "elbow" was not clearly defined, further motivating the use of complementary clustering quality metrics. The inclusion of the SC and DBI therefore proved essential, as these metrics enabled a more balanced assessment of clustering quality beyond inertia alone.

While the inertia of CL_k21 was higher than that of CL_k37 and CL_k45, both SC and DBI remained at comparable levels, indicating that CL_k21 offered a reasonable trade-off between clustering quality and computational efficiency.

### 4.2. Evaluating the results of grouping and SCE optimisation

The analysis of clustering metrics indicates that clustering based solely on hydraulic attributes yields the most physically interpretable and hydraulically consistent groupings by improving structural clarity and cluster separation within the dataset. With the exception of CHI, however, the differences were not sufficiently pronounced to support decisive conclusions.

Similarly, the analysis of WDN calibration did not reveal strongly distinguishable outcomes. In terms of pressure fit, all SCE optimisation variants achieved comparable accuracy as measured by RMSE and IOA. Nevertheless, optimisation schemes based on community detection and network coarsening consistently occupied the lower end of the performance spectrum, suggesting less robust calibration performance.

Among the k-means-based variants, CL_k21 appears as a minor outlier in terms of optimisation time, exhibiting slightly longer runtimes than CL_k27. This behaviour may indicate increased difficulty for the SCE algorithm in balancing roughness values when the number of decision variables is minimal, although the overall differences remain modest. As expected, computational time in k-means-based SCE optimisation is strongly influenced by the number of decision variables, with increasing k generally resulting in longer runtimes.

Decision variables defined by CL_HDB exhibit a different pattern. Although CL_HDB produced 40 DVs, the corresponding SCE optimisation time was substantially shorter than for the comparable CL_k37 and CL_k45 variants. This discrepancy can be attributed to the presence of the "-1" outlier group, which contained a large number of pipes but was excluded from roughness calibration, thereby reducing the effective dimensionality of the optimisation problem. In contrast, CL_HDB_NG required longer computation time than CL_k27_NG despite employing fewer clusters. This behaviour may again be related to the presence of the "-1" group; however, in this case the exclusion of a subset of pipes may have increased the complexity of the optimisation landscape.

Overall, improved clustering structure did not translate directly into improved calibration accuracy, indicating that optimisation behaviour depends not only on group separation but also on how grouping structures interact with the SCE search process.

### 4.3. Exploring the influence of attributes

ACS proved to be partially misleading in this study, particularly for the LouvainCom and VarNg variants. In both cases, high ACS values were obtained for a substantial number of attributes, despite the corresponding SCE optimisations exhibiting poor performance. An increase in the number of clusters generally led to higher ACS values for graph-based attributes, which was expected. In this respect, CL_HDB outperformed k-means, producing consistently high ACS scores across most attributes, a result that aligns with its superior SS, DBI, and CHI, as discussed earlier.

For categorical attributes such as pipe role, material, and DMA assignment, ACS values were consistently high across clustering variants, including CL_HDB. This indicates that, while CL_HDB does not exhibit a bias towards categorical attributes as strong as k-means, these attributes nonetheless play an important role in the grouping process. Similarly, flow-related attributes achieved high ACS values across most variants, confirming that flow rate was a key driver of the clustering process. Given the fundamental role of flow in determining turbulence and frictional resistance, and thus pipe roughness, this outcome aligns with first-principles hydraulic reasoning rather than being an artefact of the clustering procedure. In contrast, pipe diameter, another fundamental factor influencing turbulence, exhibited more variable ACS values, although they generally remained above **0.3**. An exception to this pattern was observed for CL_k21_NG, where both diameter- and flow-related attributes exhibited noticeably lower ACS values compared to other grouping variants. This is noteworthy given that CL_k21_NG achieved among the best clustering metrics in terms of SC and DBI, and its CHI was substantially higher than for other variants. In this context, lower ACS values indicate that the influence of individual attributes was more evenly spread across clusters, as their normalised within-cluster variance fell below the prescribed threshold less frequently. This behaviour is not inherently undesirable; the ACS does not fully capture clustering quality when the influence of individual attributes is more evenly balanced, as it is specifically designed to count instances of relatively low variance within clusters. Consequently, ACS may require revision as a clustering descriptor in future studies.

### 4.4. Identifying biases: Assessment of group size distributions

The quantitative analysis provided by Figure 6 clarifies our concerns regarding the VarNg and LouvainCom groupings. In these variants, a few large clusters dominated the SCE optimisation, which ultimately succeeded but achieved the lowest pressure-fit scores. This also explains the somewhat contradictory ACS results: although ACS values were high, suggesting a significant impact of individual attributes, the metric does not account for cluster size. As a result, a few large clusters with higher attribute variance, alongside numerous smaller clusters with lower variance, can produce a high ACS.

In CL_k27_NG, the distribution of cluster sizes is skewed, with a median around 10, indicating many small clusters. Overall, the number of pipes per cluster ranges between 5 and 110, which likely contributes to the higher CHI values observed, as CHI incorporates both between-cluster separation and within-cluster variance. A similar effect may be present for CL_HDB_NG, although its cluster-size distribution is closer to normal; the "-1" outlier cluster, containing over 300 pipes (nearly half the dataset), may substantially influence CHI, particularly as CL_HDB_NG comprises 26 clusters in total. By contrast, clusterings on the full subset (CL_k21, CL_k27, CL_k37, CL_k45, and CL_HDB) display more homogeneous cluster sizes, rarely exceeding 75 pipes per cluster, which correlates with their very similar CHI values.

Overall, this analysis justified our decision not to pursue VarNg and LouvainCom further; consequently, SCE optimisations for these variants were carried out only once instead of five times.

### 4.5. Implications of repeatability and DV boundaries

Repeated runs (Table 3) demonstrated that the stability of the SCE optimisation is not guaranteed. Several factors may contribute to this variability. Due to the embedded nature of the SCE algorithm, we could not examine the optimisation step by step. The robustness of the WDN model, its relatively low dynamics, and certain topological aspects, such as very short pipes present in the dataset, are among the most plausible explanations. Moreover, the objective of the SCE was to fit simulated pressures to measured values by adjusting pipe DVs derived from clustering. From the algorithm's perspective, roughness values serve as a means to achieve the target pressure, whereas from a practical standpoint, pressure itself was our primary goal. Inconsistencies in calibrated roughness were therefore expected.

Examining both repeatability and the frequency with which DVs reached their prescribed bounds provided key insight into optimisation performance. CL_k45, CL_HDB, and

CL_k21 achieved high overall scores; however, the RI values for CL_k21 and CL_HDB could potentially be improved by adjusting DV roughness bounds, as both variants exhibited higher boundary attainment than CL_k45. The superior performance of CL_k45 may therefore stem from identifying an optimal configuration that enables the SCE algorithm to satisfy the target function consistently across runs.

### 4.6. The elusiveness of roughness calibration

As shown in Figure 7 and Table 4, the remaining clustering variants exhibit broadly similar cluster structures, consistent with the patterns observed in the clustering metrics, ACS descriptors, and quantitative evaluations. Pipe roughness values obtained via the SCE optimisation are likewise largely consistent with those derived from manual calibration, at least for the hydraulically significant pipes examined within the analysed DMA. In terms of overall agreement between manual and heuristic calibration, the reported MAPE values are influenced by a limited number of pipes exhibiting high RAE. This indicates that global error measures are sensitive to localised discrepancies, whereas the majority of pipes show comparatively small deviations.

While CL_k45 achieved the highest repeatability, CL_HDB followed closely and required nearly half the computational time for SCE roughness optimisation (Table 3). The CL_k21 variant required a similar computation time to CL_HDB but achieved higher repeatability. Overall, these results support the credibility of the proposed heuristic; however, further testing on additional datasets and with more comprehensive pressure measurements would be necessary to strengthen confidence in its broader applicability.

## 5. Conclusions

The k-means algorithm proved to be the most suitable method for clustering the WDN in the context of roughness calibration. Increasing the number of clusters generally improved repeatability and, consequently, solution reliability, although the smallest tested k achieved comparable performance. HDBSCAN demonstrated effectiveness similar to k-means, but its lower repeatability indicates that further investigation is required before it can be considered equally robust. The selected set of hydraulic and graph-derived attributes was sufficient to produce meaningful network partitions for calibration purposes. While this observation was suggested in our previous work, the present study provides a more detailed and systematic confirmation. Notably, clustering based solely on hydraulic attributes yielded favourable clustering metrics; however, the exclusion of graph-based attributes led to less stable roughness optimisation, indicating that topological information contributes to the robustness of the SCE calibration process. These findings indicate that reliable roughness calibration depends not only on cluster compactness but also on the structural coherence of the resulting decision variables.

The embedded nature of the SCE algorithm does not provide sufficient internal information to fully explain the observed reliability issues. As a result, identifying a single dominant cause for the variability in optimisation outcomes remains challenging. This limitation is further compounded by the hydraulic conditions of the examined WDN, whose relatively low dynamics and structural characteristics hinder definitive roughness calibration. Further improvements could be achieved by incorporating additional pressure measurements to support a more comprehensive validation of the calibration results. Evaluating the methodology on additional datasets would also help to further assess its general applicability.

In this study, we examined only one of the many stages involved in roughness estimation and hydraulic model calibration. The proposed methods could therefore be further investigated and extended in several directions, including:

- Different representations of a WDN model, for example by increasing the level of skeletonization, or subsetting, either for the entire network or specifically for pipes selected for calibration.
- Alternative grouping algorithms, such as k-median or g-means clustering, which are conceptually similar to k-means and may offer comparable or improved performance.
- Different sets of attributes for grouping. While acquiring additional hydraulic attributes may be impractical, some degree of attribute reduction could be considered.
- Improved definition of DVs bounds, where introducing adaptive or data-driven boundary selection could potentially enhance optimisation robustness.
- Alternative roughness optimisation algorithms, such as neural networks, achieved comparable accuracy [11] with improved computational efficiency relative to SCE.

The proposed methodology is particularly suited to WDNs that are well described, initially calibrated, and do not require state estimation, As demonstrated by the applied metrics and descriptors, certain solutions may appear effective when assessed superficially, yet exhibit limitations when examined in greater depth.

Calibration results are presented for a representative DMA, where the SCE-optimised roughness values generally match manually calibrated values. While other DMAs were not individually validated due to increased complexity, the applied methodology ensures consistent treatment across the network, and the performance metrics indicate overall reliability.

Although the methodology presented here focuses on the calibration of a specific dataset, the extensive analysis and transparent description provided allow it to be readily adapted and repurposed for future research and practical applications.

## 6. Acknowledgment

This research was partially supported by the Ministry of Science and Higher Education, industrial PhD programme, DWD/6/0543/2022 and the Academia Profesorum Iuniorum, funded by the Wrocław University of Science and Technology.

## 7. Data availability

The research was carried out using data collected by the Municipal Water and Sewerage Company in Wrocław and was obtained under an agreement on providing data for scientific purposes. The data can only be accessed by requesting the data owner.

## 8. CRediT

- Conceptualization: Karol Dykiert (KD), Mateusz Stolarski (MS), Michał Czuba (MC), Wojciech Cieżak (WC), Piotr Bródka (PB),
- Data curation: KD,
- Formal analysis: KD,
- Funding acquisition: PB,
- Investigation: KD, MS, MC,
- Methodology: KD, MS, MC, WC, PB,
- Project administration: KD,
- Software: KD, MS, MC,
- Supervision: PB, WC,
- Validation: KD, MS, MC,
- Visualization: KD,
- Writing – original draft: KD, MS, MC,
- Writing – review and editing: KD, MS, MC, WC, PB.

# Appendix A. Dataset description

The analysed system is a high-pressure zone (HPZ) of the Wrocław water distribution network, consisting of five District Meter Areas (DMAs) supplied by a pumping station that regulates pressure. While the zone includes several physical interconnections with neighbouring DMAs, these links remain closed under normal operating conditions. The HPZ was created to meet the pressure requirements of numerous multi-storey residential buildings. Data supplied by the Municipal Water and Sewerage Company of Wrocław (MPWiK) were complemented with short-term field measurements obtained from hydrant-mounted pressure loggers installed between 11 and 24 September 2023, providing a 14-day monitoring period. Owing to practical limitations, no hydrant flow tests were performed during the model calibration. Prior to the measurement campaign, network operating conditions were verified through routine hydrant inspections, analysis of the recorded failures, and operational events in the water distribution system. Key network properties and calibrated model parameters are listed in Table A.1.

*Table A.1. Element counts in the HPZ hydraulic model. Reproduced from [19].*

| Element type | Total count |
|---|---|
| Number of DMAs | 5 |
| Number of junctions | 1247 |
| Number of pipes | 1341 |
| Number of customers (water demand points) | 1164 |
| Number of water demand patterns | 1034 |
| Number of pumps | 3 |
| Number of reservoirs | 1 |
| Pipe length in kilometres | 68,2 |
| Average total daily water demand (during 14-day period), $m^3$ | 8362 |
| Number of temporary pressure measurement stations (TPMS) | 30 |
| TPMS sampling interval, minutes | 1 |
| Number of permanent pressure measurement stations (PPMS) | 3 |
| PPMS sampling interval, minutes | 5 |

# Appendix B. Initial model calibration description

The initial model calibration consisted of following steps:

1. topology correction – assigning elevations to nodes based on a Digital Elevation Model and merging similar pipes;
2. water demand processing – creating water demand patterns and assigning them to nodes;
3. pressure data smoothing – filtering the reference pressure measurements;
4. data screening – removing nighttime measurements from all time-series data.

In summary, the dataset represents a simplified (skeletonised) HPZ model that preserves the original network structure, according to MPWiK requirements. Nodal demands are described in high detail due to the large number of individual water demand patterns. The model supports extended period simulations based on synthetic time-series data, which were developed to eliminate periods of very low flow rates. Subsequently, pipes suitable for roughness calibration were identified. These included pipes exhibiting signs of scaling, while permanently closed links were excluded, along with one DMA that was removed due to hydraulic conditions that prevented further calculations. Of the five remaining DMAs, one was sufficiently calibrated through the manual procedure, as confirmed by an industry expert; the results obtained for this DMA were therefore used as a reference for comparison.

## Appendix C. SCE hyperparameters overview

Table C.1. Shuffled Complex Evolution hyperparameters.

| Parameter | Value |
|---|---|
| Maximum no. of calls | 280000 |
| Number of complexes | 7 |
| Number of points in each complex | 65 |
| Minimum number of complexes | 5 |
| Number of evolution steps | 4 |
| Number of points in each sub-complex | 5 |
| Target objective | 17 |
| Maximum loops of convergence | 30 |
| Min. Relative change in objective function | 0.005 |

## Appendix D. Technical data

Table D.1. Implementation details: software and hardware specifications.

| Parameter | Value / version |
|---|---|
| Modelling software | DHI Mike+ 2025 Update 1 |
| Hydraulic engine | Epanet 2.2 |
| Programming environment | Python 3.12.9 |
| scikit-learn library | 1.6.1 |
| pandas library | 2.2.3 |
| numpy library | 2.1.3 |
| matplotlib library | 3.10.0 |
| NetworkX library | 3.5 |
| Operating System | Windows 11 Business |
| Processor | Intel(R) Core(TM) i7-14700 (2.10 GHz) |
| RAM | 64 GB |
| Graphic card | NVIDIA RTX 2000 Ada Generation |